# $L^2$ series solution of the relativistic Dirac-Morse problem for all energies

A. D. Alhaidari

*Physics Department, King Fahd University of Petroleum & Minerals, Box 5047, Dhahran 31261, Saudi Arabia*
Email: haidari@mailaps.org

We obtain analytic solutions for the one-dimensional Dirac equation with the Morse potential as an infinite series of square integrable functions. These solutions are for all energies, the discrete as well as the continuous. The elements of the spinor basis are written in terms of the confluent hypergeometric functions. They are chosen such that the matrix representation of the Dirac-Morse operator for continuous spectrum (i.e., for scattering energies larger than the rest mass) is tridiagonal. Consequently, the wave equation results in a three-term recursion relation for the expansion coefficients of the wavefunction. The solution of this recursion relation is obtained in terms of the continuous dual Hahn orthogonal polynomials. On the other hand, for the discrete spectrum (i.e., for bound states with energies less than the rest mass) the spinor wave functions result in a diagonal matrix representation for the Dirac-Morse Hamiltonian.

## I. INTRODUCTION

Most of the known exactly solvable nonrelativistic problems fall within distinct classes of what is referred to as "shape invariant potentials" [1]. Supersymmetric quantum mechanics, potential algebras, path integration, and point canonical transformations are four methods among many which are used in the search for exact solutions of the wave equation. These formulations were extended to other classes of conditionally exactly [2] and quasi exactly [3] solvable problems where all or, respectively, part of the energy spectrum is known. Recently, the relativistic extension of some of these developments was carried out where several relativistic problems are formulated and solved exactly [4].

In all of these developments, the objective is to find solutions of the eigenvalue wave equation $H|\chi\rangle = E|\chi\rangle$, where $H$ is the Hamiltonian and $E$ is the energy which is either discrete (for bound states) or continuous (for scattering states). The solution of this equation for a general physical system is often very difficult to obtain. However, for simple systems or for those with high degree of symmetry, an analytic solution is feasible. On the other hand, in a large class of problems that model realistic physical systems, the Hamiltonian $H$ could be written as the sum of two components: $H = H_0 + V$. The "reference" Hamiltonian $H_0$ is often simpler and carries a high degree of symmetry while the potential component $V$ is not. However, it is usually endowed with either one of two properties. Its contribution is either very small compared to $H_0$ or is limited to a finite region in configuration or function space. Perturbation techniques are used to give a numerical evaluation of its contribution in the former case, while algebraic methods are used in the latter. Thus, the analytic problem is limited to finding the solution of the reference $H_0$–problem, $(H_0 - E)|\chi\rangle = 0$. To obtain an approximate solution to the full problem in the case where the potential $V$ is short range, we could use

one of two alternative methods. In one, the contribution of $V$ is limited to a finite region in configuration space. That is, the full Hamiltonian is approximated by

$$H \cong \begin{cases} H_0 + V(r) & r < R \\ H_0 & r \geq R \end{cases}, \qquad (1.1a)$$

where $R$ is the effective range of the potential. In the other method, the matrix representation of the Hamiltonian, in a suitable basis $\{\psi_n\}_{n=0}^{\infty}$, is approximated by

$$H_{nm} \cong \begin{cases} (H_0)_{nm} + V_{nm} & n,m < N \\ (H_0)_{nm} & n,m \geq N \end{cases}, \qquad (1.1b)$$

for some adequately large integer $N$. That is, the potential is confined to a finite region in function space where it is approximated by its matrix representation in a finite subset of the basis, $\{\psi_n\}_{n=0}^{N-1}$. This latter approach to the solution is called the *J*-matrix method [5]. It is endowed with formal and computational similarities to the former approach– the celebrated *R*-matrix method [6]. They both use square integrable basis functions to carry out the calculation.

The utilization of $L^2$ methods in scattering calculations underwent several major developments starting with the "stabilization technique" of Hazi and Taylor [7]. It was then followed by the development of the $L^2$-Fredholm method [8] in which scattering information is extracted from the discrete eigenvalues of the full Hamiltonian and those of the reference Hamiltonian in a finite $L^2$-basis $\{\psi_n\}_{n=0}^{N-1}$. This method is analogous to discretizing the continuous spectrum of the system by confining it to a box in configuration space. The *J*-matrix method was the next substantial development in this formulation in which the contribution of the reference Hamiltonian is taken fully into account. On the other hand, there is a limited class of short range potentials where a full analytic solution to the problem is possible. In nonrelativistic quantum mechanics one such class that has a symmetry which is associated with the dynamical group SO(2,1) includes the Morse, Pöschl-Teller, Eckart, etc. potentials. Due to the higher degree of symmetry of such problems, it is frequently possible to find a special basis for the solution space that could support a tridiagonal matrix representation for the wave operator $H - E$. This property makes the solution of the problem easily attainable as will be explained next. Let $\{\psi_n\}_{n=0}^{\infty}$ be such a basis, which is complete, square integrable and compatible with the domain of the Hamiltonian. Therefore, the action of the wave operator on the elements of the basis takes the general form $(H - E)|\psi_n\rangle \sim |\psi_n\rangle + |\psi_{n-1}\rangle + |\psi_{n+1}\rangle$, where we can also write

$$\langle \psi_n | H - E | \psi_m \rangle = (a_n - y)\delta_{n,m} + b_n \delta_{n,m-1} + b_{n-1}\delta_{n,m+1}, \qquad (1.2)$$

where $y$ and the coefficients $\{a_n, b_n\}_{n=0}^{\infty}$ are real and, in general, functions of the energy $E$, the angular momentum $\ell$, and the potential parameters. Therefore, the matrix representation of the wave equation, which is obtained by expanding $|\chi\rangle$ as $\sum_m f_m |\psi_m\rangle$ in $(H - E)|\chi\rangle = 0$ and projecting on the left by $\langle \psi_n |$, results in the following three-term recursion relation

$$y f_n = a_n f_n + b_{n-1} f_{n-1} + b_n f_{n+1} \qquad (1.3)$$

Consequently, the problem translates into finding solutions of the recursion relation for the expansion coefficients of the wavefunction. In most cases this recursion is solved easily and directly by correspondence with those for well known orthogonal polynomials. An example of a problem which is already solved using this approach is the non-relativistic Morse problem which was tackled successfully by J. T. Broad [9] and P. C. Ojha [10]. The nonrelativistic Coulomb problem, despite being long range, was also solved by Yamani and Reinhardt using the same approach in which the expansion coefficients of the wavefunction are written in terms of the Pollaczek polynomials [11]. The relativistic extension of the Coulomb problem has recently been carried out by this author where the expansion coefficients of the spinor wavefunction are written in terms of the Meixner-Pollaczek polynomials [12]. In this article the same approach will also be employed in finding exact analytic solution of the relativistic Dirac-Morse problem for all energies.

It is obvious that the solution of (1.3) is obtained modulo an overall factor which is a function of *y* but, otherwise, independent of *n*. The uniqueness of the solution is achieved by the requirement that, for example, the wavefunction be energy normalizable. It should also be noted that the solution of the problem as given by Eq. (1.3) above is obtained for all *E*, the discrete as well as the continuous, constrained only by the reality and boundedness of the representation. Moreover, the matrix representation of the wave equation, Eq. (1.2), clearly shows that the discrete spectrum is easily obtained by diagonalization which requires that:

$$b_n = 0, \text{ and } a_n - y = 0 \qquad (1.4)$$

In Sec. II, we set up the one dimensional Dirac equation for a two-component spinor coupled non-minimally to the potential $(A_0, A_1)$. The space and time components of the potential are taken to be of the same type. A global unitary transformation is applied to the Dirac equation to separate the variables such that the resulting second order differential equation for the spinor components becomes Schrödinger-like. This results in a simple and straightforward correspondence between the relativistic and nonrelativistic problems. The correspondence will be used in Sec. III as a guide for constructing a square integrable basis for the solution space of the Dirac-Morse problem. In this construction we impose the requirement that the matrix representation of the Dirac operator be tridiagonal. The result is a three-term recursion relation for the expansion coefficients of the spinor wavefunction. The recursion relation is written in a form that makes its solution easily attainable by simple and direct comparison with that of the continuous dual Hahn orthogonal polynomials [13]. We conclude with a short discussion in Sec. IV.

## II. THE DIRAC-MORSE PROBLEM

The one-dimensional Dirac equation for a free particle reads $\left(i\hbar\gamma^\mu\partial_\mu - mc\right)\chi = 0$, where *m* is the rest mass of the particle, *c* the speed of light, and $\chi$ is the two-component spinor wavefunction. The summation convention over repeated indices is used. That is, $\gamma^\mu\partial_\mu = \gamma^0\partial_0 + \gamma^1\partial_1 = \gamma^0\frac{\partial}{c\partial t} + \gamma^1\frac{\partial}{\partial x}$, where $\gamma^0$ and $\gamma^1$ are two constant square matrices satisfying the anticommutation relation $\{\gamma^\mu, \gamma^\nu\} = \gamma^\mu\gamma^\nu + \gamma^\nu\gamma^\mu = 2\mathcal{G}^{\mu\nu}$ and $\mathcal{G}$ is the space-time metric which is equal to diag$(+,-)$. A two-dimensional matrix representation that satisfies this relation is chosen as $\gamma^0 = \sigma_3 = \begin{pmatrix} 1 & 0 \\ 0 & -1 \end{pmatrix}$ and $\gamma^1 = i\sigma_2 = \begin{pmatrix} 0 & 1 \\ -1 & 0 \end{pmatrix}$. In the atomic units $\hbar = m = 1$, the Compton wavelength, $\lambdabar = \hbar/mc = 1/c$, is the relativistic parameter and Dirac equation reads $\left(i\gamma^\mu \partial_\mu - \lambdabar^{-1}\right)\chi = 0$. Next, we let the Dirac spinor be coupled to the two component potential $A_\mu = (A_0, A_1)$. Gauge invariant coupling, which is accomplished by the "minimal" substitution $\partial_\mu \to \partial_\mu + i\lambdabar A_\mu$, transforms the free Dirac equation into

$$\left[i\gamma^\mu(\partial_\mu + i\lambdabar A_\mu) - \lambdabar^{-1}\right]\chi = 0, \tag{2.1}$$

which, when written in details, reads as follows

$$i\frac{\partial}{\partial t}\chi = \left(-i\lambdabar^{-1}\sigma_1\frac{\partial}{\partial x} + \sigma_1 A_1 + A_0 + \lambdabar^{-2}\sigma_3\right)\chi, \tag{2.2}$$

where $\sigma_1 = \begin{pmatrix} 0 & 1 \\ 1 & 0 \end{pmatrix}$. For time independent potentials, this equation gives the following matrix representation of the Dirac Hamiltonian (in units of $mc^2 = 1/\lambdabar^2$)

$$H = \begin{pmatrix} 1 + \lambdabar^2 A_0 & \lambdabar^2 A_1 - i\lambdabar\frac{d}{dx} \\ \lambdabar^2 A_1 - i\lambdabar\frac{d}{dx} & -1 + \lambdabar^2 A_0 \end{pmatrix} \tag{2.3}$$

Thus the eigenvalue wave equation reads $(H - \varepsilon)\chi = 0$, where $\varepsilon$ is the relativistic energy which is real, dimensionless and measured in units of $1/\lambdabar^2$.

Equation (2.1) is invariant under the local gauge transformation

$$A_\mu \to A_\mu + \partial_\mu \Lambda, \quad \chi \to e^{-i\lambdabar\Lambda}\chi, \tag{2.4}$$

where $\Lambda(t,x)$ is a real space-time scalar function. Consequently, the off diagonal term $\lambdabar^2 A_1$ in the Hamiltonian (2.3) could be eliminated ("gauged away") by a suitable choice of the gauge field $\Lambda(x)$ in the transformation (2.4). However, our choice of coupling will be non-minimal, which is obtained by the replacement $\lambdabar^2 A_1 \to \pm i\lambdabar^2 A_1$, respectively. That is the Hamiltonian (2.3) is replaced by the following

$$H = \begin{pmatrix} 1 + \lambdabar^2 A_0 & i\lambdabar^2 A_1 - i\lambdabar\frac{d}{dx} \\ -i\lambdabar^2 A_1 - i\lambdabar\frac{d}{dx} & -1 + \lambdabar^2 A_0 \end{pmatrix} \tag{2.5}$$

Next, we take the two components of the potential, $A_0$ and $A_1$, to be of the same functional form by writing $A_0 = V(x)$ and $A_1 = \frac{1}{\lambdabar\xi}V(x)$, where $\xi$ is a real positive parameter with inverse length dimension. Writing $\chi$ as $\begin{pmatrix} ig(x) \\ f(x) \end{pmatrix}$ gives us the following two-component wave equation

$$\begin{pmatrix} 1 + \lambdabar^2 V - \varepsilon & \lambdabar\left(\frac{V}{\xi} - \frac{d}{dx}\right) \\ \lambdabar\left(\frac{V}{\xi} + \frac{d}{dx}\right) & -1 + \lambdabar^2 V - \varepsilon \end{pmatrix}\begin{pmatrix} g \\ f \end{pmatrix} = 0 \tag{2.6}$$

This matrix equation results in two coupled first order differential equations for the two spinor components. Eliminating one component in favor of the other gives a second order differential equation. This will not be Schrödinger-like. That is, it contains first order derivatives. Obtaining a Schrödinger-like wave equation is desirable because it results in a substantial reduction of the efforts needed for getting the solution. It puts at our disposal a variety of well established techniques to be employed in the analysis and solution of the problem. These techniques have been well developed over the years by many researchers in dealing with the Sturm-Liouville problem and the Schrödinger equation. One such advantage, which will become clear shortly, is the resulting map between the parameters of the relativistic and nonrelativistic problems. This parameter map could be used in obtaining, for example, the relativistic energy spectrum in a simple and straight-forward manner from the known nonrelativistic spectrum.

To obtain a Schrödinger-like equation we apply a global unitary transformation $\mathcal{U}(\eta) = \exp(\frac{i}{2}\lambdabar\eta\sigma_2)$ to the Dirac equation (2.6), where $\eta$ is a real constant parameter and $\sigma_2 = \begin{pmatrix} 0 & -i \\ i & 0 \end{pmatrix}$. The Schrödinger-like requirement results in the constraint that $\sin(\lambdabar\eta) = \pm\lambdabar\xi$, where $-\frac{\pi}{2} < \lambdabar\eta < +\frac{\pi}{2}$. It is to be noted that the angular parameter of the unitary transformation was intentionally split as $\lambdabar\eta$ and not collected into a single angle, say $\varphi$. This is suggested by investigating the constraint $\sin(\varphi) = \pm\lambdabar\xi$ in the nonrelativistic limit ($\lambdabar \to 0$) where we should have $\sin(\varphi) \approx \varphi = \pm\lambdabar\xi$. It also makes it obvious that in the nonrelativistic limit the transformation becomes the identity (i.e., not needed). The unitary transformation together with the parameter constraint map Eq. (2.6) into the following one

$$\begin{pmatrix} C - \varepsilon + (1\pm 1)\lambdabar^2 V & \lambdabar\left(\mp\xi + \frac{C}{\xi}V - \frac{d}{dx}\right) \\ \lambdabar\left(\mp\xi + \frac{C}{\xi}V + \frac{d}{dx}\right) & -C - \varepsilon + (1\mp 1)\lambdabar^2 V \end{pmatrix} \begin{pmatrix} \phi^+(x) \\ \phi^-(x) \end{pmatrix} = 0, \quad (2.7)$$

where $C = \cos(\lambdabar\eta) = \sqrt{1 - (\lambdabar\xi)^2} > 0$ and

$$\begin{pmatrix} \phi^+ \\ \phi^- \end{pmatrix} = \mathcal{U}\psi = \begin{pmatrix} \cos\frac{\lambdabar\eta}{2} & \sin\frac{\lambdabar\eta}{2} \\ -\sin\frac{\lambdabar\eta}{2} & \cos\frac{\lambdabar\eta}{2} \end{pmatrix} \begin{pmatrix} g \\ f \end{pmatrix} \quad (2.8)$$

Equation (2.7) gives the following ("kinetic balance") relation between the two spinor components

$$\phi^{\mp}(x) = \frac{\lambdabar}{C \pm \varepsilon}\left[-\xi \pm \frac{C}{\xi}V(x) + \frac{d}{dx}\right]\phi^{\pm}(x) \quad (2.9)$$

While, the resulting Schrödinger-like wave equation becomes

$$\left[-\frac{d^2}{dx^2} + \left(\frac{C}{\xi}\right)^2 V^2 \mp \frac{C}{\xi}\frac{dV}{dx} + 2\varepsilon V - \frac{\varepsilon^2 - 1}{\lambdabar^2}\right]\phi^{\pm}(x) = 0 \quad (2.10)$$

This equation could be rewritten in a form which is more familiar in the language of supersymmetric quantum mechanics

$$\left[-\frac{d^2}{dx^2} + \left(W^2 \mp \frac{dW}{dx}\right) - \frac{(\varepsilon/C)^2 - 1}{\lambdabar^2}\right]\phi^{\pm}(x) = 0, \quad (2.11)$$

where $W(x) = \frac{C}{\xi}V(x) + \frac{\xi}{C}\varepsilon$. $W^2 \pm W'$ are two superpartner potentials sharing the same energy spectrum (i.e., they are "isospectral") except for the highest positive energy state and the lowest negative energy state, where $\varepsilon = \pm C$, respectively. These two states belong only to $W^2 - W'$ [14].

The nonrelativistic limit ($\lambdabar \to 0$) gives $\varepsilon \approx 1 + \lambdabar^2 E$ and $C \approx 1 - \frac{1}{2}\lambdabar^2\xi^2$. Therefore, Eq. (2.9) shows that $\phi^+$ is the larger of the two relativistic spinor components (i.e., $\phi^+$ is the component that survives the nonrelativistic limit, whereas $\phi^- \sim \lambdabar \phi^+ \to 0$). Consequently, if we favor the upper spinor component then our choice of sign in the transformation parameter constraint is the top + sign. That is, we choose $\sin(\lambdabar \eta) = +\lambdabar \xi$.

For the Dirac-Morse problem the potential function $V(x) = -Ae^{-\omega x}$, where the parameters $A$ and $\omega$ are real and $\omega > 0$. Substituting this potential into Eq. (2.10) gives the following Schrödinger-like second order differential equation for the upper spinor component

$$\left[ -\frac{d^2}{dx^2} + \left(\frac{CA}{\xi}\right)^2 e^{-2\omega x} - \frac{CA}{\xi}(\omega + 2\xi\varepsilon/C)e^{-\omega x} - \frac{\varepsilon^2 - 1}{\lambdabar^2} \right] \phi^+(x) = 0 \qquad (2.12)$$

We compare this equation with that of the nonrelativistic Schrödinger-Morse problem

$$\left[ -\frac{d^2}{dx^2} + B^2 e^{-2\omega x} - B(\omega + 2D)e^{-\omega x} - 2E \right] \Phi(x) = 0, \qquad (2.13)$$

where $B$ and $D$ are real, and $B > 0$. The comparison gives the following correspondence between the parameters of the two problems:

$$E \to (\varepsilon^2 - 1)/2\lambdabar^2, \; B \to \pm CA/\xi, \; D \to \begin{cases} \xi\varepsilon/C \\ -\omega - \xi\varepsilon/C \end{cases} \qquad (2.14)$$

The top (bottom) choice of map for $B$ and $D$ corresponds to positive (negative) values of $A$. Using this parameter map and the well-known nonrelativistic energy spectrum [15], $E_n = -\frac{1}{2}\omega^2 \left(\frac{D}{\omega} - n\right)^2$, we obtain the following bound states relativistic spectrum

$$\frac{\varepsilon_n^\pm}{C} = \begin{cases} \lambdabar^2 \xi \omega n \pm \sqrt{1 - (\lambdabar C\omega n)^2} & , \; A > 0 \\ -\lambdabar^2 \xi \omega(n+1) \pm \sqrt{1 - [\lambdabar C\omega(n+1)]^2} & , \; A < 0 \end{cases} \qquad (2.15)$$

where $n = 0, 1, 2, ..., n_{max}$ for $A > 0$, $n = 0, 1, 2, ..., n_{max} - 1$ for $A < 0$, and $n_{max}$ is the maximum integer which is less than or equal to $1/\lambdabar C\omega$. The two energy spectra are related by $\varepsilon_n^\pm \big|_{A<0} = -\varepsilon_{n+1}^\mp \big|_{A>0}$. Therefore, all states are degenerate except two. Those are the highest positive, and lowest negative energy states corresponding to $\varepsilon_0^\pm \big|_{A>0} = \pm C$, respectively. In fact, these are the two states that belong only to the super-potential $W^2 - W'$ in Eq. (2.11) above.

The upper spinor component could be obtained from the nonrelativistic wave-function [16],

$$\Phi_n(x) \sim z^{\alpha_n} e^{-z/2} L_n^{2\alpha_n - 1}(z), \qquad (2.17)$$

using the same parameter map (2.14), where $z = \frac{2B}{\omega} e^{-\omega x}$, $\alpha_n = \frac{D}{\omega} - n$, and $L_n^\lambda(z)$ is the Laguerre polynomial [17]. Square integrability requires that $\alpha_n > 0$ (i.e., $n \leq D/\omega$).

The above findings, which are valid only for bound states, will be used in the following section as a guide to writing down an $L^2$ spinor bases for the solution space of scattering states. These bases will be chosen such that the matrix representation of the Dirac-Morse operator $(H - \varepsilon)$ in Eq. (2.7) is tridiagonal.

## III. TRIDIAGONAL REPRESENTATIONS

As suggested by the expression of the nonrelativistic wavefunction in (2.17) and the parameter map (2.14), we write $z = \pm \frac{2CA}{\omega\xi} e^{-\omega x} = \frac{2C|A|}{\omega\xi} e^{-\omega x}$ for $\pm A > 0$. This maps the whole real line into its positive half. That is, $x \in [+\infty, -\infty] \to z \in [0, +\infty]$. In this coordinate notation and with the choice $\sin(\lambda\eta) = +\lambda\xi$, the Dirac-Morse equation (2.7) reads as follows:

$$\begin{pmatrix} C - \varepsilon \mp \lambda^2(\omega\xi/C)z & -\lambda\omega\left(\frac{\xi}{\omega} \pm \frac{z}{2} - z\frac{d}{dz}\right) \\ -\lambda\omega\left(\frac{\xi}{\omega} \pm \frac{z}{2} + z\frac{d}{dz}\right) & -C - \varepsilon \end{pmatrix} \begin{pmatrix} \phi \\ \theta \end{pmatrix} = 0, \quad (3.1)$$

where $\begin{pmatrix} \phi \\ \theta \end{pmatrix} = \begin{pmatrix} \phi^+ \\ \phi^- \end{pmatrix}$, and the top (bottom) sign corresponds to positive (negative) value of $A$. A square integrable basis function (with respect to the measure $dx = -dz/z\omega$) in this configuration space which is compatible with the domain of the wave operator and satisfies the boundary conditions could be written as

$$\phi_n = \sqrt{\frac{\omega\Gamma(n+1)}{\Gamma(n+\nu+1)}} (\beta z)^\alpha e^{-\beta z/2} L_n^\nu(\beta z), \quad (3.2)$$

where $\alpha$ and $\beta$ are real, positive, dimensionless parameters and $\nu > -1$. The non-relativistic wavefunction (2.17) and the parameter map (2.14) support this choice of basis. The "kinetic balance" relation (2.9) in the new coordinate notation reads as follows

$$\theta = \frac{-\lambda\omega}{C+\varepsilon}\left(\frac{\xi}{\omega} \pm \frac{z}{2} + z\frac{d}{dz}\right)\phi, \quad (3.3)$$

for $\pm A > 0$. It suggests that the lower component of the spinor basis is obtainable from the upper by the following general differential relation

$$\theta_n \sim \lambda\left(\mu + \zeta z + z\frac{d}{dz}\right)\phi_n, \quad (3.4)$$

where the dimensionless parameters $\mu$ and $\zeta$ are real and will be determined as we proceed. Substituting (3.2) into (3.4) and using the differential and recursion properties of the Laguerre polynomials (shown in the Appendix) we obtain

$$\theta_n = -\lambda\tau\sqrt{\frac{\omega\Gamma(n+1)}{\Gamma(n+\nu+1)}}(\beta z)^\alpha e^{-\beta z/2}\left\{2\left[\beta\left(\mu + \alpha - \frac{\nu+1}{2}\right) + 2\zeta\left(n + \frac{\nu+1}{2}\right)\right]L_n^\nu(\beta z)\right.$$
$$\left. -(\beta + 2\zeta)(n+\nu)L_{n-1}^\nu(\beta z) + (\beta - 2\zeta)(n+1)L_{n+1}^\nu(\beta z)\right\} = -2\lambda\beta\tau\left(\mu + \zeta z + z\frac{d}{dz}\right)\phi_n \quad (3.5)$$

where $\tau$ is another real parameter. In this spinor basis $\left\{\psi_n = \begin{pmatrix}\phi_n \\ \theta_n\end{pmatrix}\right\}_{n=0}^\infty$, the matrix representation of the Dirac-Morse operator of (3.1) reads

$$\langle\psi_n|H-\varepsilon|\psi_m\rangle = (C-\varepsilon)\langle\phi_n|\phi_m\rangle \mp \lambda^2(\omega\xi/C)\langle\phi_n|z|\phi_m\rangle - \left(C+\varepsilon - \frac{\omega}{\beta\tau}\right)\langle\theta_n|\theta_m\rangle$$
$$+ \lambda(\mu\omega - \xi)\left[\langle\theta_n|\phi_m\rangle + \langle\theta_m|\phi_n\rangle\right] + \lambda\omega(\zeta \mp 1/2)\left[\langle\theta_n|z|\phi_m\rangle + \langle\theta_m|z|\phi_n\rangle\right] \quad (3.6)$$

where we have used integration by parts in writing $\langle\phi_n|z\overrightarrow{\frac{d}{dz}}|\theta_m\rangle = -\langle\phi_n|\overleftarrow{\frac{d}{dz}}z|\theta_m\rangle$ since the product $\phi_n(z)\theta_m(z)$ vanishes at the boundaries $z = 0$ and $z \to \infty$, and

$$\langle\phi_n|z\overrightarrow{\frac{d}{dz}}|\theta_m\rangle = \int_{-\infty}^{+\infty}\phi_n z\frac{d\theta_m}{dz}dx = \frac{1}{\omega}\int_0^\infty \phi_n\frac{d\theta_m}{dz}dz$$
$$= \frac{-1}{\omega}\int_0^\infty \theta_m\frac{d\phi_n}{dz}dz = -\int_{-\infty}^{+\infty}\theta_m z\frac{d\phi_n}{dz}dx = -\langle\phi_n|\overleftarrow{\frac{d}{dz}}z|\theta_m\rangle \quad (3.7)$$

Now, we require that the representation (3.6) be tridiagonal. That is, $\langle \psi_n | H - \varepsilon | \psi_m \rangle = 0$ for all $|n - m| \geq 2$. For the first two terms on the right side of Eq. (3.6) to comply with this requirement we must have $2\alpha = \nu + 1$. Moreover, the two terms inside the last square brackets on the right side of the equation destroy the tridiagonal structure. Thus, the multiplying factor must vanish. That is, we must choose $\zeta = \pm 1/2$ for $\pm A > 0$. In addition, the term $\langle \theta_n | \theta_m \rangle$ results in a tridiagonal representation only if $(\beta - 1)(\beta + 1) = 0$, otherwise the multiplying factor must vanish. The latter choice is not acceptable since it requires that $\tau = \frac{\omega/\beta}{C+\varepsilon}$ which violates the "kinetic balance" relation (3.3) that will be necessary later on for further investigations. Now, since $\beta$ must be positive, then we end up with the only possibility that $\beta = 1$. Consequently, we are left with three undetermined real parameters ($\alpha$, $\mu$, and $\tau$) and with the following two components of the spinor basis

$$\phi_n = \sqrt{\frac{\omega \Gamma(n+1)}{\Gamma(n+2\alpha)}} (z)^\alpha e^{-z/2} L_n^{2\alpha - 1}(z) \tag{3.8a}$$

$$\theta_n = -2\hbar \tau \sqrt{\frac{\omega \Gamma(n+1)}{\Gamma(n+2\alpha)}} (z)^\alpha e^{-z/2} \begin{cases} (\mu + \alpha) L_n^{2\alpha - 1}(z) - z L_{n-1}^{2\alpha}(z) &, A > 0 \\ (\mu + \alpha) L_n^{2\alpha - 1}(z) - z L_n^{2\alpha}(z) &, A < 0 \end{cases} \tag{3.8b}$$

Substituting these into (3.6) and using the orthogonality and recurrence relations of the Laguerre polynomials (shown in the Appendix) we obtain, after some manipulations, the following elements of the symmetric tridiagonal matrix representation of the Dirac-Morse operator

$$(H - \varepsilon)_{n,n} = C - \varepsilon \mp 2\hbar^2 (\omega \xi/C)(n + \alpha) \pm 4\hbar^2 \tau(\xi - \mu\omega)(n + \alpha \pm \mu)$$
$$- 4\hbar^2 \tau^2 (C + \varepsilon - \omega/\tau) \left[ 2n^2 + 2n(2\alpha \pm \mu \mp 1/2) + (\alpha \pm \mu)^2 + (1 \mp 1)\alpha \right] \tag{3.9a}$$

$$(H - \varepsilon)_{n,n-1} = \hbar^2 \sqrt{n(n + 2\alpha - 1)} \times$$
$$\left[ \pm (\omega \xi/C) \pm 2\tau(\mu\omega - \xi) + 4\tau^2 (C + \varepsilon - \omega/\tau)\left(n + \alpha \pm \mu - \tfrac{1 \pm 1}{2}\right) \right] \tag{3.9b}$$

where the top and bottom signs correspond to $\pm A > 0$, respectively. If we define the following quantities:

$$p(\varepsilon) = 4\tau^2 (C + \varepsilon - \omega/\tau) \text{ and } q = 2\tau(\mu\omega - \xi), \tag{3.10}$$

then the matrix representation of the wave equation $(H - \varepsilon)|\chi\rangle = 0$, where $|\chi\rangle = \sum_m f_m |\psi_m\rangle$, results in the following three-term recursion relation for the expansion coefficients of the wavefunction

$$\left[ 2n^2 + 2n(2\alpha \pm \mu \mp 1/2) + (\alpha \pm \mu)^2 \pm 2(n + \alpha)\tfrac{q + \omega\xi/C}{p} + (1 \mp 1)\alpha \pm 2\tfrac{\mu q}{p} + \tfrac{\varepsilon - C}{\hbar^2 p} \right] f_n$$
$$- \sqrt{n(n + 2\alpha - 1)} \left( n - \tfrac{1 \pm 1}{2} + \alpha \pm \mu \pm \tfrac{q + \omega\xi/C}{p} \right) f_{n-1} \tag{3.11}$$
$$- \sqrt{(n+1)(n + 2\alpha)} \left( n + \tfrac{1 \mp 1}{2} + \alpha \pm \mu \pm \tfrac{q + \omega\xi/C}{p} \right) f_{n+1} = 0$$

for $n \geq 1$. Rewriting (3.11) in terms of the polynomial $Q_n(\varepsilon) = \sqrt{\Gamma(n+1)/\Gamma(n+2\alpha)} f_n(\varepsilon)$, gives the following recursion relation

$$\left[ 2n^2 + 2n(2\alpha \pm \mu \mp 1/2) + (\alpha \pm \mu)^2 \pm 2(n + \alpha)\tfrac{q + \omega\xi/C}{p} \pm 2\tfrac{\mu q}{p} + (1 \mp 1)\alpha + \tfrac{\varepsilon - C}{\hbar^2 p} \right] Q_n$$
$$- n\left( n - \tfrac{1 \pm 1}{2} + \alpha \pm \mu \pm \tfrac{q + \omega\xi/C}{p} \right) Q_{n-1} - (n + 2\alpha)\left( n + \tfrac{1 \mp 1}{2} + \alpha \pm \mu \pm \tfrac{q + \omega\xi/C}{p} \right) Q_{n+1} = 0 \tag{3.12}$$

for $\pm A > 0$. We compare this with the recursion relation satisfied by the continuous dual Hahn orthogonal polynomials $S_n^\lambda(y^2;a,b)$ [13] that reads

$$y^2 S_n^\lambda = \left[(n+\lambda+a)(n+\lambda+b) + n(n+a+b-1) - \lambda^2\right] S_n^\lambda$$
$$-n(n+a+b-1)S_{n-1}^\lambda - (n+\lambda+a)(n+\lambda+b)S_{n+1}^\lambda \qquad (3.13)$$

where $y^2 > 0$. The comparison results in the following values for the parameters of the polynomial

$$\lambda = a = \alpha, \begin{cases} b = \mu + \frac{q+\omega\xi/C}{p} &, \quad y^2 = \frac{C-\varepsilon}{\lambda^2 p} - \mu(\mu + 2q/p) \; ; \; A > 0 \\ b = 1 - \mu - \frac{q+\omega\xi/C}{p} &, \quad y^2 = \frac{C-\varepsilon}{\lambda^2 p} - \mu(\mu - 2q/p) \; ; \; A < 0 \end{cases} \qquad (3.14)$$

Therefore, we can write

$$f_n(\varepsilon) = \sqrt{\frac{\Gamma(n+2\alpha)}{\Gamma(n+1)}} \, S_n^\alpha(y^2;\alpha,b), \qquad (3.15)$$

which is defined up to a multiplicative factor that depends on $\varepsilon$ but, otherwise, independent of $n$. The continuous dual Hahn polynomials could be written in terms of the hypergeometric function ${}_3F_2$ as, [13]

$$S_n^\lambda(y^2;a,b) = {}_3F_2\left(\begin{matrix}-n,\lambda+iy,\lambda-iy\\ \lambda+a,\lambda+b\end{matrix}\bigg|1\right) \qquad (3.16)$$

The orthogonality relation associated with these polynomials is as follows

$$\int_0^\infty \rho^\lambda(y) S_n^\lambda(y^2;a,b) S_m^\lambda(y^2;a,b) dy = \frac{\Gamma(n+1)\Gamma(n+a+b)}{\Gamma(n+\lambda+a)\Gamma(n+\lambda+b)} \delta_{nm}, \qquad (3.17)$$

where $\rho^\lambda(y) = \frac{1}{2\pi}\left|\frac{\Gamma(\lambda+iy)\Gamma(a+iy)\Gamma(b+iy)}{\Gamma(\lambda+a)\Gamma(\lambda+b)\Gamma(2iy)}\right|^2$. Therefore, the normalizable $L^2$ series solution of the Dirac-Morse problem could be written as

$$\chi(x,\varepsilon) = N^\alpha(\varepsilon) \sum_{n=0}^\infty \sqrt{\frac{\Gamma(n+2\alpha)}{\Gamma(n+1)}} \, S_n^\alpha\left(y(\varepsilon)^2;\alpha,b(\varepsilon)\right) \psi_n(x), \qquad (3.18)$$

where $N^\alpha(\varepsilon) = \sqrt{\rho^\alpha(y)(dy/d\varepsilon)}$ is a normalization factor that makes $\chi$ energy-normalizable, whereas the two components of the spinor basis element $\psi_n(x)$ are those given by Eqs. (3.8).

Further analysis of these solutions, such as obtaining the discrete spectrum, is tractable only if the "kinetic balance" relation (3.3) is strictly imposed on the basis elements. That is, relation (3.4) should be identical to (3.3) which requires that

$$\mu = \xi/\omega \text{ and } \tau = \frac{\omega/2}{C+\varepsilon}, \qquad (3.19)$$

where $\varepsilon \neq -C$. This gives $q = 0$ and $p = \frac{-\omega^2}{C+\varepsilon}$ resulting in the following polynomial parameters: $\lambda = a = \alpha$, $y(\varepsilon)^2 = (\varepsilon^2 - 1)/\lambda^2\omega^2$ and $b(\varepsilon) = \begin{cases} -(\xi/C\omega)\varepsilon &, \; A>0 \\ 1+(\xi/C\omega)\varepsilon &, \; A<0 \end{cases}$. Thus, the energy dependence of the parameters simplifies resulting in the following expression for the wavefunction

$$\chi(x,\varepsilon) = N^\alpha(\varepsilon) \sum_{n=0}^\infty \sqrt{\frac{\Gamma(n+2\alpha)}{\Gamma(n+1)}} \, S_n^\alpha\left(\frac{\varepsilon^2-1}{\lambda^2\omega^2};\alpha,\frac{1\mp 1}{2} \mp \frac{\xi\varepsilon}{C\omega}\right) \psi_n(x), \qquad (3.20)$$

for $\pm A > 0$. If we take the nonrelativistic limit ($\lambda \to 0$, $\varepsilon \approx 1 + \lambda^2 E$, and $C \approx 1 - \frac{1}{2}\lambda^2\xi^2$) of this solution we obtain

$$\chi(x,E) = N_{NR}^\alpha(E) \sum_{n=0}^{\infty} \sqrt{\frac{\Gamma(n+2\alpha)}{\Gamma(n+1)}} \, S_n^\alpha\left(\frac{2E}{\omega^2};\alpha,-\frac{D}{\omega}\right) \phi_n(x), \tag{3.21}$$

which agrees with the findings by J. T. Broad [9] and P. C. Ojha [10].

The requirement that the argument $y^2$ of $S_n^a(y^2;b,c)$ be positive implies that the solution (3.20) is valid for $|\varepsilon|>1$. In other words, the solution (3.18) or (3.20) is for energies larger than the rest mass $mc^2$ corresponding to scattering states. Solutions for $|\varepsilon|<1$ are for bound states and correspond to $y^2<0$. To obtain these solutions, which are different from (3.20), and to calculate the discrete energy spectrum, we impose the diagonalization requirement (1.4). In the case of the recursion relation (3.11) and with the parameters assignment (3.19), this requirement translates into the following conditions

$$n+\alpha+b=0, \text{ and } y^2=-\alpha^2, \tag{3.22}$$

giving exactly the same energy spectrum in Eq. (2.15). Therefore, both the parameter correspondence map method (presented near the end of Sec. II) and the diagonalization method have independently given the same results. In addition, we also obtain the following as a consequence of the diagonalization requirement (3.22):

$$\alpha_n = \begin{cases} (\xi/C\omega)\varepsilon_n - n & , \quad A>0 \\ -(\xi/C\omega)\varepsilon_n - n - 1 & , \quad A<0 \end{cases} \tag{3.23}$$

where $n \leq \xi/C\omega$ and $n \leq \xi/C\omega - 1$ for $\pm A > 0$, respectively. The bound states spinor wavefunctions become

$$\phi_n = \sqrt{\frac{\omega\Gamma(n+1)}{\Gamma(n+2\alpha_n)}} (z)^{\alpha_n} e^{-z/2} L_n^{2\alpha_n-1}(z) \tag{3.24a}$$

$$\theta_n = -\frac{\hbar\omega}{C+\varepsilon_n} \sqrt{\frac{\omega\Gamma(n+1)}{\Gamma(n+2\alpha_n)}} (z)^{\alpha_n} e^{-z/2} \begin{cases} (\alpha_n+\xi/\omega)L_n^{2\alpha_n-1}(z) - zL_{n-1}^{2\alpha_n}(z) & , \quad A>0 \\ (\alpha_n+\xi/\omega)L_n^{2\alpha_n-1}(z) - zL_n^{2\alpha_n}(z) & , \quad A<0 \end{cases} \tag{3.24b}$$

## IV. DISCUSSION

We would like to conclude with some comments that have to do with the type and number of solutions of the recursion relation (3.11) or (3.12). Typically, there are two solutions to such three-term recursion relation. This could be understood by noting that the orthogonal polynomials that satisfy the recursion relation are at the same time solutions of a second order differential equation. In other words, there is a correspondence between three-term recursion relations and second order differential equations for a given set of initial relations or boundary conditions, respectively. The solutions obtained above in (3.18) could be termed "regular solutions." These correspond to solutions of the recursion relation in terms of polynomials of the "first kind". Polynomials of the "second kind" satisfy the same recursion relation (for $n \geq 1$) but with a different initial relation (for $n = 0$). These correspond to "irregular solutions", or in a more precise term "regularized solutions," since they are regular at the origin of configuration space while behaving asymptotically as the irregular solution.

For a large positive integer $N$ the recursion relation (3.12) could be rewritten as $2uQ_{N+n}(u) - Q_{N+n-1}(u) - Q_{N+n+1}(u) = 0$, where $u = 1 - \frac{1}{2}\left(\frac{y}{N}\right)^2$. Defining $\hat{Q}_n(u) \equiv Q_{N+n}(u)$, we could write it as

$$2u\hat{Q}_n(z) - \hat{Q}_{n-1}(u) - \hat{Q}_{n+1}(u) = 0 \tag{4.1}$$

This is the recursion relation for the Chebyshev polynomials. For large $n$, they are oscillatory (i.e., they behave like sine's and cosine's). The two independent oscillatory solutions of (4.1), which will be designated by $Q_n^\pm(u)$, differ by a phase. The origin of this phase difference could be traced back to the initial relation ($n = 0$) of the recursion (3.12). Thus, the initial relation must have two different forms. This difference propagates through the recursion to the asymptotic solutions. One of these initial relations is homogeneous and corresponds to the regular solution, which was obtained above. The other is inhomogeneous and corresponds to the regularized solution. They could be written as:

$$\left[y^2 - \alpha(\alpha + 2b)\right]Q_0^+ + 2\alpha(\alpha + b)Q_1^+ = 0 \tag{4.2a}$$

$$\left[y^2 - \alpha(\alpha + 2b)\right]Q_0^- + 2\alpha(\alpha + b)Q_1^- = \mathcal{W} \neq 0, \tag{4.2b}$$

where $\mathcal{W} = \mathcal{W}(y^2, b)$ and is related to the Wronskian of the two solutions. For scattering problems, the phase shift is obtained by the analysis of the two solutions $Q_n^\pm(u)$. Such analysis is typical of algebraic scattering methods in nonrelativistic quantum mechanics. A clear example is found in the *J*-matrix method of scattering [5,10].

## ACKNOWLEDGMENT

The author is grateful to M.E.H. Ismail for the help in identifying the solution of the three-term recursion relation (3.12) as the continuous dual Hahn orthogonal polynomials.

## APPENDIX: PROPERTIES OF THE LAGUERRE POLYNOMIALS

The following are useful formulas and relations satisfied by the generalized orthogonal Laguerre polynomials $L_n^\nu(x)$ that are relevant to the developments carried out in this work. They are found in most textbooks on orthogonal polynomials [17]. We list them here for ease of reference.

The differential equation:
$$\left[x\frac{d^2}{dx^2} + (\nu + 1 - x)\frac{d}{dx} + n\right]L_n^\nu(x) = 0, \tag{A.1}$$
where $\nu > -1$ and $n = 0, 1, 2, \ldots$

Expression in terms of the confluent hypergeometric function:
$$L_n^\nu(x) = \frac{\Gamma(n+\nu+1)}{\Gamma(n+1)\Gamma(\nu+1)} {}_1F_1(-n; \nu + 1; x) \tag{A.2}$$

The three-term recursion relation:
$$xL_n^\nu = (2n + \nu + 1)L_n^\nu - (n + \nu)L_{n-1}^\nu - (n+1)L_{n+1}^\nu \tag{A.3}$$

Other recurrence relations:
$$xL_n^\nu = (n + \nu)L_n^{\nu-1} - (n+1)L_{n+1}^{\nu-1} \tag{A.4}$$

$$L_n^\nu = L_n^{\nu+1} - L_{n-1}^{\nu+1} \tag{A.5}$$

Differential formula:

$$x\frac{d}{dx}L_n^\nu = nL_n^\nu - (n+\nu)L_{n-1}^\nu \tag{A.6}$$

Orthogonality relation:

$$\int_0^\infty \rho^\nu(x) L_n^\nu(x) L_m^\nu(x) dx = \frac{\Gamma(n+\nu+1)}{\Gamma(n+1)} \delta_{nm}, \tag{A.7}$$

where $\rho^\nu(x) = x^\nu e^{-x}$.